\documentclass{aa}
\usepackage{graphicx}
\usepackage{epsfig}
\usepackage{natbib}
\bibpunct{(}{)}{;}{a}{}{,}
\errorcontextlines = \maxdimen
\newcommand{\dphi}{(\Delta\phi)}

\newcommand{\NPulses}{2\,320}
\newcommand{\NPulse}{5\,200}
\newcommand{\mr}{\mathrm}

\mathcode`\;="013B
\newdimen\breedte
\begin{document}

\title{On the Search for Coherent Radiation from Radio Pulsars}  

\author{J.M. Smits \inst{1} \and B.W. Stappers \inst{2,3} \and J-P. Macquart \inst{4}\and
R. Ramachandran \inst{2,3} \and J. Kuijpers \inst{1}}

\institute{Department of Astrophysics, University of Nijmegen, 
Nijmegen \and Stichting `Astron', PO Box 2, 7990 AA Dwingeloo \and Astronomical
Institute `Anton Pannekoek', Amsterdam \and Kapteyn Institute,
University of Groningen, Groningen}

\offprints{J.M. Smits \email{roysm@astro.kun.nl}}

\date{Received <date> / Accepted <date>}

\abstract{
We have examined data from pulsars B0950+08 and B0329+54 for evidence of
temporally coherent radiation using the modified coherence function (MCF) technique
of~\citet{Jenet}. We consider the influence of both instrumental bandpass and interstellar propagation effects. 
Even after removal of the effects due to the instrumental bandpass, we detect a
signature in the MCF of our PSR B0329+54 data 
which is consistent with the definition of a coherent signal.
However, we model the effects due to interstellar scintillation for this pulsar and show
that it reproduces the observed signature. In particular, the temporal coherence time is close to the
reciprocal of the decorrelation bandwidth due to diffractive scintillation.
Furthermore, comparison of the coherence times of three pulsars reported by
~\citet{Jenet} with their expected diffractive decorrelation
bandwidths suggests that the detection of coherence in these pulsars is
also likely a result of interstellar scintillation, and is not intrinsic to
the pulsars.
}

\maketitle

\section{Introduction}
Only a few months after the discovery of the first pulsars in
1967~\citep{Hewish} it was realised
they were in fact neutron stars~\citep{Gold, Pacini}. However, the emission process, responsible
for the radio-frequency radiation in pulsars, remains uncertain. One
characteristic of the pulsar radiation is a very high brightness
temperature which can reach $10^{30}$\,K~\citep[e.g.][]{Manchester}, which implies that it is generated by a coherent
mechanism. It is therefore important to identify characteristics of coherent emission in pulsar radiation.
\citet{Rickett75} suggested that the observed radiation is well described by
amplitude modulated Gaussian noise, which contains no
information about the coherent nature of the
radiation. \citet{Cordes76} later generalized this to non-Gaussian shot
  noise. \citet{Hankins03} have seen individual shot nano-pulses in
the giant pulses from the Crab pulsar, suggesting that the intrinsic
structure is less than 2 ns.

\citet{Jenet} (hereafter JAP) claim to have detected the existence of
coherent non-Gaussian radiation on 100 ns time
  scales for pulsars B0823+26, B0950+08 and
B1133+16 in observations made at the Arecibo observatory. They used their
``modified coherence function'' (MCF), defined 
in Eq.~(\ref{eq:MCFr}) below, to find statistics inconsistent with amplitude-modulated Gaussian noise in the voltage 
time series from observations of these three pulsars. They show that a 
coherent model for pulsar radiation can account for the observed statistics.
However, if the MCF of a time series of pulsar radiation shows statistics of
a non-Gaussian nature, it does not yet prove that the pulsar radiation itself
contains non-Gaussian statistics. We have to consider the effect
of the ISM on pulsar radiation and the instrumental effects
on the signal after detection. 
JAP assert that scintillation does not influence the statistics of the
signal measured by the MCF. In particular, they state that if the statistics of the
intensity fluctuations are well described by Gaussian noise, the MCF
ought not to exceed zero as a result of propagation effects.
Scintillation, however,
can give rise to a quasi-periodic fluctuation in the frequency power
spectrum (see Fig.~\ref{fig:Bandpass}b). Such fluctuations in the
frequency domain can be expected to influence the autocorrelations from
which the MCF is constructed.

We consider here an independent data set for pulsars
B0950+08 and B0329+54 to try to confirm the detection of coherence. In Sect.~\ref{sec:observations}
we describe our observations and the construction
of the MCF. In Sect.~\ref{sec:results} we present an analysis of our
data and investigate whether scintillation and instrumental effects
do influence the MCF. Here, we also present a numerical
experiment that shows that scintillation quantitatively explains the
coherent features observed by JAP and compare their values of the
coherence time with the inverse of the diffractive decorrelation bandwidth for each pulsar. In Sect.~\ref{sec:discussion} we discuss our results and present our conclusions.

\section{Observations and analysis}
\label{sec:observations}
For our analysis, we use data from pulsars B0329+54 and
B0950+08. These are bright pulsars and it is therefore possible to
detect single pulses with the high signal-to-noise ratio required to find a clear signature of coherent
radiation. 

These data were taken using the Westerbork Synthesis Radio Telescopes (WSRT) with its
pulsar backend, PuMa~\citep{Voute}. In its tied array mode the WSRT is equivalent to
a single dish with a diameter of 94\,m and has a gain of
1.2\,K/Jy.
PSR B0950+08 was observed on 28 April 1999 at a centre frequency of 382\,MHz,
with a bandwidth of 10\,MHz. PSR B0329+54 was observed
on 13 August and 11 September 1999 at a centre frequency of
328\,MHz, with a bandwidth of 5\,MHz. In both observations a 10\,MHz
band was Nyquist-sampled in 2 linear polarisation channels. For PSR
B0329+54 the 5\,MHz band was formed by digitally filtering the data
using a finite impulse response filter. After sampling, the data were
2 bit digitised. The high time resolution of 50\,ns
and 100\,ns for pulsars B0950+08 and B0329+54, respectively, is
necessary to study radiation which is expected to contain features
with a coherence time of a few hundred nanoseconds~\citep{Jenet}. 

We remove the effects of
interstellar dispersion by means of coherent
dedispersion~\citep{Hankins}. Finally, for both pulsars, we take two
sets of small successive time windows for each individual pulse. 
One set contains small windows centered on the peak of the
average pulse profile,
the other set contains a region outside the pulse, which is used for
system plus sky noise corrections.
\begin{table}
\flushleft
\caption{Pulsar Characteristics and Analysis Parameters}
  \begin{tabular}{lcc}
	\hline
	Pulsar			& B0950+08	& B0329+54 	\\
	\hline
	Period			& 0.2531\,s	& 0.7145\,s	\\
	Period derivative	& $8.6\times 10^{-17}$ & $2.04959\times10^{-15}$\\
	Dispersion measure	&  2.9702\,pc\,cm$^{-3}$ & 26.7765\,pc\,cm$^{-3}$  \\
	Centre frequency	& 382\,MHz	& 328\,MHz	\\
	Bandwidth		& 10\,MHz	& 5\,MHz	\\
	Number of Pulses	& \NPulses	& \NPulse	\\
	Size of window		& 102.4\,$\mu$s	& 204.8\,$\mu$s	\\
	Number of windows	& 16		& 16		\\
	Time-resolution			& 50\,ns	& 100\,ns 	\\
	\hline
  \end{tabular}
\label{tab:parameters}
\end{table}
The size, number of windows, and other parameters can be found in
Table~\ref{tab:parameters}. To make sure that the periodicity of
microstructure does not show up in our results as coherence, we chose a window 
smaller than the typical microstructure time-scale for both pulsars, which is 0.17\,ms
for PSR B0950+08~\citep{Rickett} and $0.6 - 1.5$\,ms for PSR B0329+54~\citep{Lange98}. 

From these windows we calculate the MCF for real voltages. The MCF
tests the fourth moment of the signal against the square of the
second moment and is given by
\begin{equation}
  \label{eq:MCFr}
  \mathrm{M}_S\dphi\equiv\frac{\langle C_{I_S}\dphi\rangle}{\langle
  C_{I_S}(0)\rangle}-\frac{1}{3}\left(2\left [\frac{\langle
  C_S\dphi\rangle}{\langle C_S(0)\rangle}\right ]^2+1\right),
\end{equation}
where $S$ is the time series of real voltages, $I_S=S\cdot S$ is
the intensity, while $C_S(\Delta\phi)$ and $C_{I_S}(\Delta\phi)$ are the
autocorrelations of $S$ and $I_S$, respectively. The angular brackets denote
averaging over different windows and different pulses. In the analysis of JAP, $S$ is a
time series of complex voltages. Although the MCF for complex voltages is different 
from the MCF for real voltages, \emph {physically}, they are the same.
\begin{figure*}[htb]
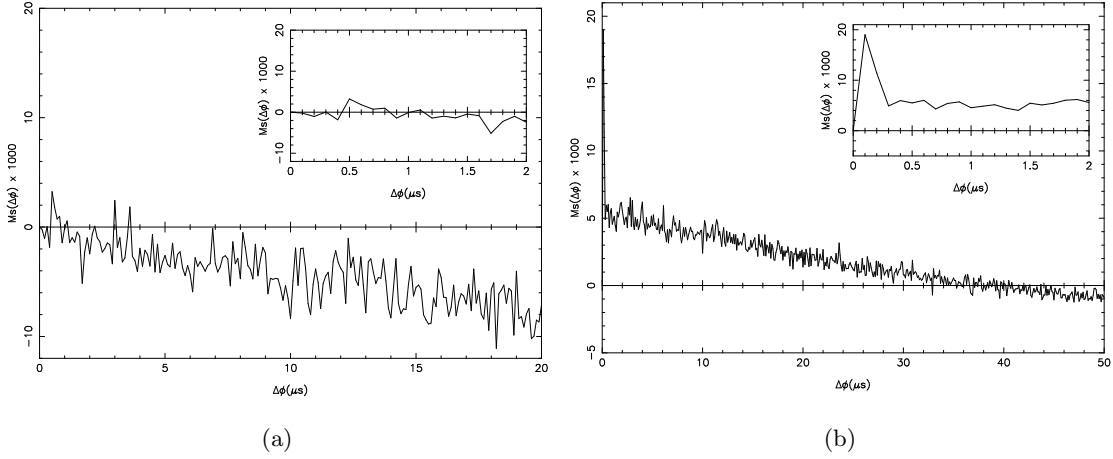

  \centering
  \begin{tabular}{c@{\hspace{8pt}}c}
  \breedte=0.41\textwidth
  \includegraphics[angle=-90, width=0.4\textwidth]{H4303F1.PS}
	&  \includegraphics[angle=-90, width=0.4\textwidth]{H4303F2.PS}\\
		& \\
  (a)		& (b)\\
  \end{tabular}
  \caption{Our results for the MCF as a function of phase delay calculated for pulsars B0950+08 (a)
  and B0329+54 (b) before bandpass correction, obtained from \NPulses~and \NPulse~time series of
  real data obtained with WSRT, respectively. The time resolution is 50\,ns for PSR B0950+08
  and 100\,ns for PSR B0329+54. Note that the
  peak at zero phase delay has been set to zero.}
  \label{fig:OurResults1}
\end{figure*}

The MCF is sensitive to phase
  relationships between measurement points separated in time, called temporal coherence.
The two polarisations are also separated and treated
as different pulses. We thus obtain 32 time series from each
pulse.

\subsection{Calculating the MCF}
The MCF is sensitive to coherent features present in any time series.
For pulsar radiation, such features could result when the radiation is
emitted in bunches of coherently radiating particles~\citep[e.g.][]{Ruderman}. However, if
there is extensive \emph{incoherent} addition of these bunches, the
resulting radiation will contain only Gaussian statistics~\citep{Cordes76}. 
Because of the way the MCF is constructed (see Eq.~(\ref{eq:MCFr})), it
has a value of zero for Gaussian noise (if applied to complex data) and falls below zero with
increasing delay for amplitude modulated Gaussian noise.
Therefore, if the
MCF of a time series becomes larger than zero, it contains statistics
of a non-Gaussian nature, which \emph{might} give information on
bunching of coherently radiating particles.
To find such statistics, we wish to calculate the MCF for a noise-subtracted
pulsar signal.

In order to calculate the second term of the MCF we note that the recorded on-pulse voltage time series
$V(t)$ contains both the pulsar signal and
system plus sky noise. It can therefore be expressed as
\begin{equation}
\label{eq:Vt}
V(t)=S(t)+N(t),
\end{equation}
where $S(t)$ is the pulsar signal and $N(t)$ is the system plus
sky noise, which is contained in the off-pulse signal.
Using the voltage time series of the on- and off-pulse, the average autocorrelation (AAC) of the pulsar
signal and the AAC of the intensity of the pulsar signal are
calculated as follows.
First, one window containing one polarisation of the on-pulse of
the pulsar profile is Fourier transformed to the
frequency domain. The value at each frequency is multiplied by its
complex conjugate and
inverse Fourier transformed back into the time domain (correlation theorem).
This results in the temporal autocorrelation of one window of on-pulse signal. The
autocorrelations of all the windows in the total time series of
on-pulse signal (containing both polarisations) are then averaged
together to give the AAC of the on-pulse, $\langle
  C_V\dphi\rangle$. The same calculation is made for the off-pulse
signal to give $\langle C_N\dphi\rangle$. The AAC
of the pulsar signal without the system and sky noise, is given by
\begin{equation}
\label{eq:Cs}
\langle C_S\dphi\rangle = \langle C_V\dphi\rangle - \langle
C_N\dphi\rangle. 
\end{equation}
This AAC is divided by the average pulsar
intensity of the total time series, which is simply the average
intensity of the on-pulse minus the average intensity
of the off-pulse,
\begin{equation}
\label{eq:Is}
\langle I_S\rangle = \langle I_V\rangle - \langle
I_N\rangle.
\end{equation}
Squaring the result, we obtain the second term of the MCF,
$\langle C_S\dphi\rangle^2/\langle C_S(0)\rangle^2$,
where $\langle C_S(0)\rangle=\langle I_S\rangle$.

To obtain the first term, we calculate the AAC's of the intensity of the voltage
time series of on-pulse signal and off-pulse signal. The AAC of the
intensity of the noise-subtracted pulsar signal is given by
\setlength\arraycolsep{1pt}
\begin{eqnarray}
\label{eq:CIs}
\langle C_{I_S}\dphi\rangle & = & \langle C_{I_V}\dphi\rangle -
\langle C_{I_N}\dphi\rangle \nonumber \\
& & - 2\left[\langle C_V(0)\rangle-\langle
C_N(0)\rangle\right]\cdot\langle C_N(0)\rangle \nonumber \\
& & - 4\left[\langle C_V\dphi\rangle-\langle C_N\dphi\rangle\right]\cdot\langle
C_N\dphi\rangle.
\end{eqnarray}
The first term in the MCF results by dividing this AAC by the average of the square of the
intensity of the total time series, which is given by
\begin{eqnarray}
\label{eq:CIs0}
\langle C_{I_S}(0)\rangle & = & \langle C_{I_V}(0)\rangle - \langle
C_{I_N}(0)\rangle \nonumber \\
& & - 6\left[\langle C_V(0)\rangle-\langle
C_N(0)\rangle\right]\cdot\langle C_N(0)\rangle.
\end{eqnarray}
The MCF is then calculated according to Eq.~(\ref{eq:MCFr}).

\section{Results}
\label{sec:results}
Our MCF for PSR B0950+08 is shown in Fig.~\ref{fig:OurResults1}a. 
The MCF starts at a value of zero and falls with increasing $\Delta\phi$, which is due to the non-zero slope of the
average pulse profile of the pulsar.
Our MCF for PSR B0329+54 is shown in Fig.~\ref{fig:OurResults1}b. It 
shows that there is a broad excess reaching a $\Delta\phi$
of 35\,$\mu$s. There is
also a large peak extending over the second and third bins.

The peak in PSR 0329+54 is at a $\Delta\phi$ that is the reciprocal of
the bandwidth. This makes us
suspect that the frequency-dependent gain of the system, the bandpass, can influence the MCF. 
We can understand the presence of a peak in the MCF due to the bandpass as
follows\footnote{To make a clear distinction between time and frequency, we
use time ($t$) instead of phase ($\phi$).}. Define $\tilde{S}(\nu)$ as the Fourier transform of the
voltage signal $S(t)$
and define the bandpass as 
$\tilde{S}(\nu)\tilde{S}^*(\nu)$. The bandpass contains 
modulation with a width equal to the width
of the bandpass, $\Delta\nu_\mr{bp}$ and also modulation with a width
of about one third the width of the bandpass (see Figs.~\ref{fig:Bandpass}c and d). This implies that $S(t)$ and
$S(t)^2$ contain a peak with a width on the order of
$3/\Delta\nu_\mr{bp}$. The autocorrelation of the signal
$C_S(\Delta t)$ and the
autocorrelation of the intensity $C_{I_S}(\Delta t)$ will then contain
a peak at $\Delta t=0$ and decorrelate on a timescale on the order of
$3/\Delta\nu_\mr{bp}$. The MCF contains
the peak from the autocorrelation of the intensity minus 2/3 the
square of the peak of the autocorrelation of the signal itself. This
last contribution works as a partial bandpass correction, but not as
a full bandpass correction, as the two contributions are not
necessarily equal.

\subsection{Correcting for the bandpass}
The noise power is dominated by the sky background radiation at the
frequencies used in this work. We therefore take the shape of the power spectrum of 
the signal in the off-pulse as an estimate of the bandpass. The WSRT
bandpass contains modulation with a width on the order
  of 2 MHz and interference peaks~\citep[see
also][ chapter~4]{Marco}). Fig.~\ref{fig:Bandpass} shows the bandpass of
on-pulse and off-pulse emission for one polarisation state of pulsars B0950+08 and B0329+54, averaged over 2,000
pulses.

To correct for the bandpass, we calculate and apply a bandpass
correction as follows. First, we estimate the average
bandpass as the power of the off-pulse signal as a function of
frequency averaged over all pulses.
We then normalise by dividing each frequency value by the average of the bandpass.
Before the calculation of the MCF, we Fourier transform the signal
inside each window into the frequency domain, divide the value at
each frequency by the square root of the corresponding value of
the bandpass and Fourier transform back into the time domain. This
process does not correct for a possible frequency dependent phase
shift introduced by the system.
We carry out the procedure described above separately for both
polarisations. High-level interference peaks are replaced in the frequency domain by the average of their
left and right neighbours. Adjacent peaks are considered as one.
 Remaining interference peaks should only have a small influence on
 the MCF and only on long timescales.
\begin{figure}[htb]
  \centering
  \breedte=0.41\textwidth
  \includegraphics[angle=-90, width=0.4\textwidth]{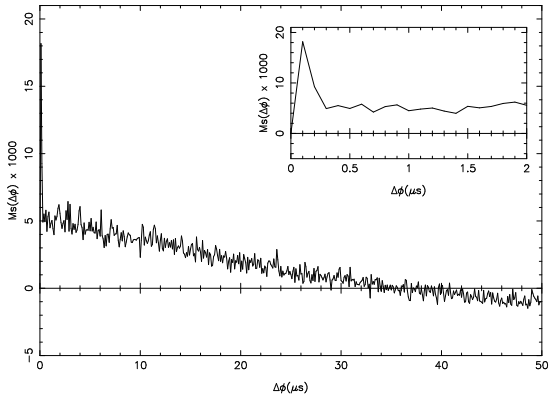}
  \caption{Our results of the MCF calculated for PSR B0329+54 after bandpass
  correction. See Fig.~\ref{fig:OurResults1} for explanation.}
  \label{fig:OurResults2}
\end{figure}

In Fig~\ref{fig:OurResults2} we see the
corrected MCF for PSR B0329+54. The peak at low $\Delta\phi$ has gone down slightly,
but remains present. Looking at Fig.~\ref{fig:Bandpass}b, we see that there is
strong modulation in the on-pulse bandpass of PSR B0329+54 which is absent
in the off-pulse bandpass. This reduces the effectiveness of the
bandpass correction, as this modulation makes the on-pulse bandpass
different from the off-pulse bandpass. We did not do a bandpass correction for PSR B0950+08, 
as there is no sign of coherence present in our data from this pulsar.

\subsection{Effect of scintillation on the MCF}

Having established that variations in the bandpass shape can theoretically influence the
MCF, we now consider the additional influence of interstellar
scintillation (ISS).  Diffractive ISS introduces structure in the instantaneous spectrum
of a pulsar and can therefore, in principle, influence the MCF.  In
particular, we argue that ISS induces a signal in the MCF that is likely
to masquerade as a false detection of temporally coherent pulsar
radiation. 

Diffractive interstellar scintillation is observed in pulsars,
including B0950+08 and B0329+54, at centimetre wavelengths and lower.  It
imposes large ($>100$\%) temporal and spectral modulations in the
intensity of the radiation.  The decorrelation timescale and bandwidth of
the fluctuations are chiefly determined by two parameters: the diffractive
scale $r_{\rm diff}$, and the Fresnel scale $r_{\rm F}$.  It is often
convenient to consider the phase fluctuations imposed by the scattering
medium confined to a thin phase screen of distance $L$ from the observer. 
Then the diffractive scale is the distance on the phase screen over which
the root mean square phase difference is one radian, and the Fresnel scale
is set by the distance to the screen and the wavelength $r_{\rm
F}=\sqrt{\lambda L/2\pi}$. 

Diffractive scintillation imposes random spectral variations of
characteristic bandwidth
\begin{eqnarray} 
\Delta \nu_{\rm dc} = \nu \left( \frac{r_{\rm diff}}{r_{\rm F}}\right)^2 
\end{eqnarray} 
across the pulsar signal, and
this spectrum changes on a time scale $\sim r_{\rm diff}/v$, where $v$ is
the velocity of the scintillation pattern across the Earth.  The expected
decorrelation bandwidth for observations of B0329+54 and B0950+08 at
their observing frequency are 29\,kHz~\citep{Cordes86} and 0.22\,GHz~\citep{Phillips92}, respectively.
These values were scaled from a frequency of 1\,GHz and 51\,MHz, respectively, assuming $\Delta \nu_{\rm
dc}\ \propto \nu^{4.4}$~\citep{Cordes85}.

Let us consider the effect of diffractive spectral variations on the quantity
$C_I(\Delta t)$.  The instantaneous power spectrum of the pulsar signal
contains large variations with a characteristic bandwidth $\Delta \nu_{\rm
dc}$. Thus, since the amplitudes of the Fourier transformed voltages
contain ripples, the
observed voltages are broadened. In the case of scintillation,
  the broadening can be described by the pulse broadening
  function (PBF)~\citep{Williamson72}. The instantanous PBF is expected to
  have wiggles on a time scale set by the reciprocal
of the decorrelation bandwidth, $\Delta t \approx 1/\Delta \nu_{\rm
  dc}$.
This implies that the intensity also contains variations on the same
characteristic time scale.  Neglecting any intrinsic temporal coherence
due to the pulsar, the intensity autocorrelation function $C_I(\Delta t)$
is expected to peak at $\Delta t=0$ and decorrelate on a timescale set by
the inverse of the scintillation decorrelation bandwidth. The average
intensity autocorrelation function, obtained by combining the
autocorrelations from many individual pulses and even over many
diffractive timescales, is expected to exhibit the same decorrelation
timescale.  This is because every set of data is
expected to exhibit spectral structure with a similar decorrelation
bandwidth. 

The second contribution to the MCF comes from the autocorrelation of the
signal voltages, $C_S(\Delta t)$.  The autocorrelation function of the
pulsar voltages, $S(t)$, and the on-pulse bandpass, $\tilde S(\nu) \tilde
S(\nu)^*$ are related by a Fourier transform: 
\begin{eqnarray}
C_S(\Delta t) \rightleftharpoons \langle \tilde{S}(\nu) \tilde{S}^*(\nu) 
\rangle, \label{eq:Cst}
\end{eqnarray}
where the angular brackets denote an average over many scintles in
either/both the temporal and spectral domains.  Thus the autocorrelation
$C_S(\Delta t)$ is the inverse Fourier transform of the {\it mean}
bandpass.  However, the mean bandpass contains little structure due to the
stochastic scintillation.  Observations over many scintillation time
scales wash out the frequency dependent structure due to scintillation, so
that the mean on-pulse bandpass reflects mainly the instrumental response to
the intrinsic spectrum of the pulsar. Thus scintillation has little effect
on $C_S(\Delta t)$. 

\begin{figure*}
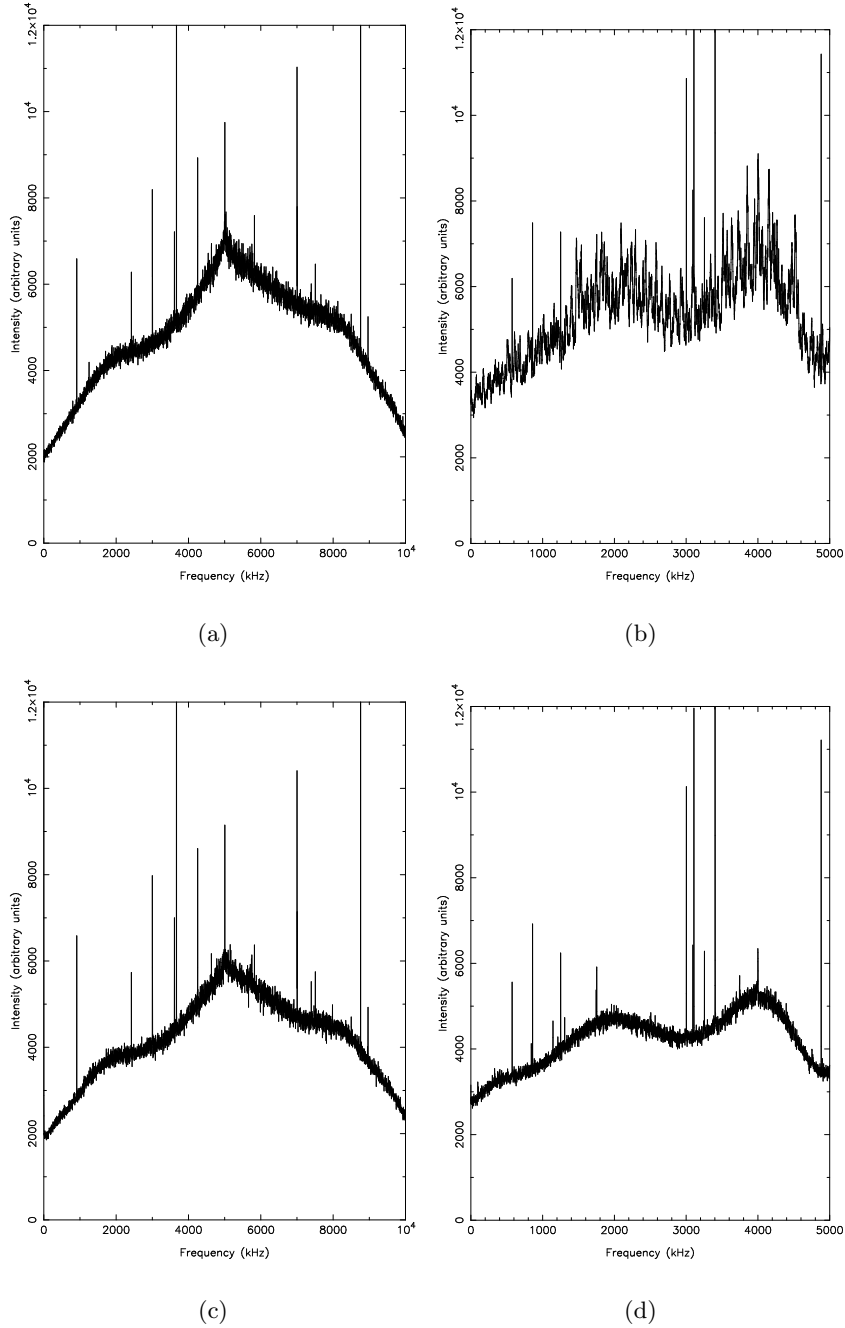

\centering
  \begin{tabular}{c@{\hspace{8pt}}c} 
    \centering
    \epsfig{file=H4303F4.PS, width=0.3\textwidth} & \epsfig{file=H4303F5.PS, width=0.3\textwidth} \\
        & \\
    (a) & (b) \\
        & \\
    \epsfig{file=H4303F6.PS, width=0.3\textwidth} & \epsfig{file=H4303F7.PS, width=0.3\textwidth} \\
	& \\
    (c) & (d) \\
        & \\
  \end{tabular}

  \caption{Bandpass of one polarisation of WSRT for (a) On-pulse of
  B0950+08, (b) On-pulse of B0329+54, (c) Off-pulse of B0950+08, (d)
  Off-pulse of B0329+54, all averaged over 2,000 pulses. The peaks are
  caused by interference. The on-pulse of B0329+54 shows scintillation
  with a width of approximately 30\,kHz (or 25 bins).}  
  \label{fig:Bandpass}
\end{figure*}
\subsection{The Modified Coherence Function due to Scintillation}
A set of synthetic observations from data generated by a scattering
simulation is constructed in order to demonstrate the effect of
scintillation on the MCF suggested by the preceeding arguments.

The simulations are conducted by constructing a set of phase
fluctuations as a function of position and wavelength, $\phi({\bf
r};\lambda)$, frozen onto a thin phase-changing screen located a
distance $L$ from the observer's plane. We take a plane wave of unit amplitude
incident on the phase screen, so that the phase of the wave upon exiting the
screen is $\phi({\bf r})$. The phase fluctuations are generated
according to a von Karman power spectrum~\citep{Narayan}
\begin{equation} 
Q_\phi( {\bf q})=Q_0\, \left( \left\vert {\bf q} \right\vert^2+q_\mr{min}^2
\right)^{-\beta/2}\exp(-| {\bf q}|/q_\mr{max}), 
\label{eq:Qq}
\end{equation} 
where $Q_0$ is the amplitude of the power spectrum and sets the
diffractive scale length, $r_{\rm diff}$.  The Kolmogorov value of
$\beta=11/3$ is assumed for the power law index~\citep{Lee}, and
$q_\mr{max}$ and $q_\mr{min}$ are the upper and lower cutoffs of the power
spectrum respectively. Each specific realisation of the scattering screen
is generated by randomly choosing the phase associated with each spectral
mode~\citep[e.g.][]{Narayan}, so that the
Fourier transform of the phase fluctuations, $\phi({\bf x} )$, is given by
\begin{eqnarray} 
\tilde \phi({\bf q}) = Q_\phi({\bf q})^{1/2} \exp[ i |{\bf q}| \cdot {X}], 
\end{eqnarray} 
with $X \in (0,2 \pi]$.  The wavefield
at the observer's plane is the inverse Fourier transform of the
Fourier transform of the wavefield at the
exit plane of the phase screen, $\exp[i \phi({\bf r})]$ multiplied by the
Fourier transform of the propagator $\exp[-i(2 \pi r_{\rm F})^2 q^2/2]$. 

\begin{figure}[!ht]
  \centering
   \epsfig{file=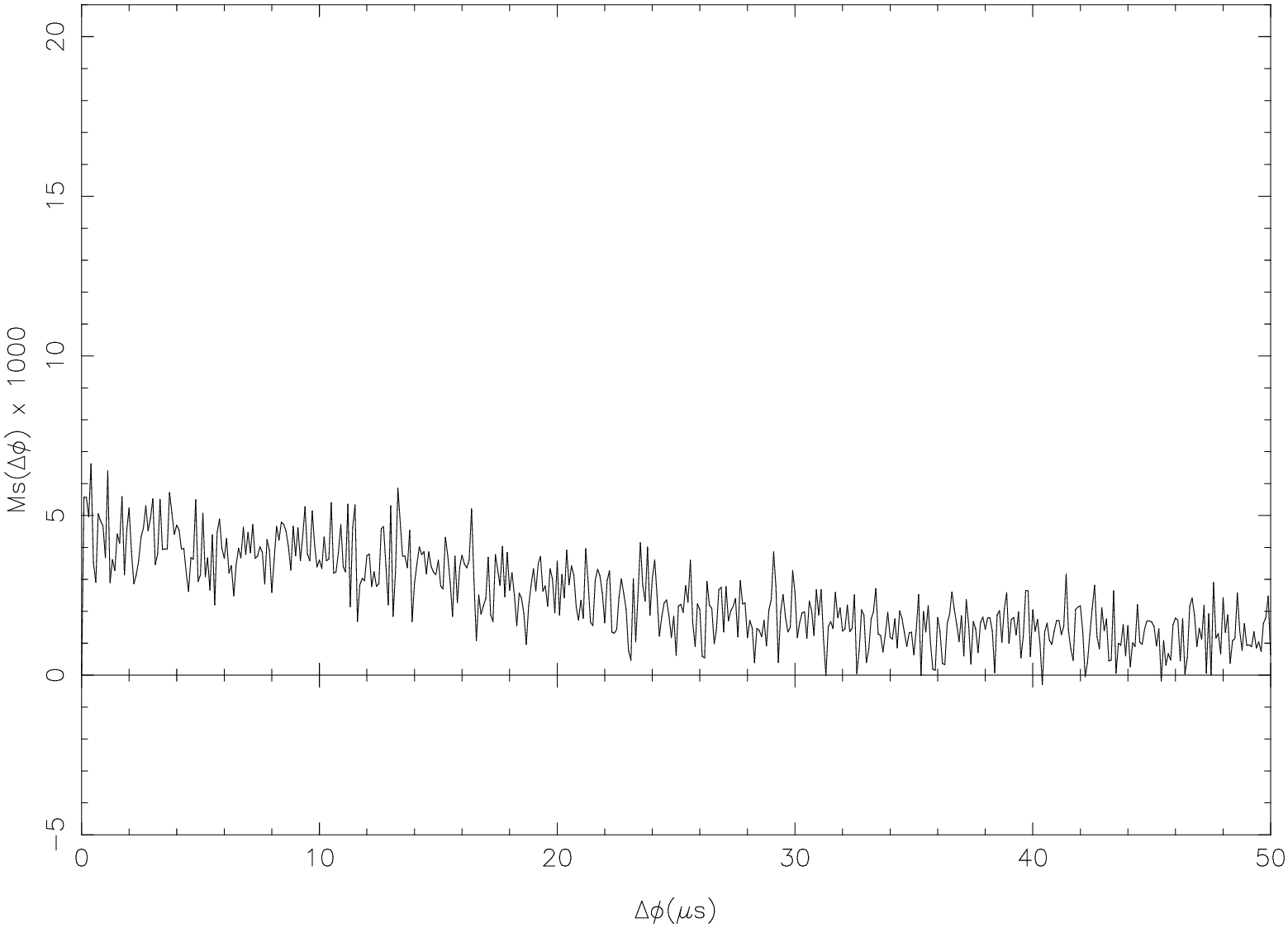, width=0.35\textwidth}
   \caption{The MCF of simulated scintillation at a frequency of 328\,MHz and
  a bandwidth of 5\,MHz. The input parameters are set to generate
  scintillation similar to that in our data from B0329+54}
  \label{fig:SimScint}
\end{figure}

Temporal fluctuations in the wavefield are obtained by moving the phase screen 
relative to the observer, however, due to the limited number of grid points 
short time scale variations could not be simulated. Spectral variations are obtained by scaling 
the phase fluctuations according to 
\begin{eqnarray}
\phi({\bf r};\lambda) = \frac{\lambda}{\lambda_0} \phi({\bf r};\lambda_0),
\end{eqnarray}
and by altering the Fresnel scale in a similar manner, 
$r_{\rm F}(\lambda)  = \sqrt{\lambda/\lambda_0} r_{\rm F}(\lambda_0)$  
where $\lambda$ is a wavelength within the observation bandwidth
and $\lambda_0$ is an arbitrary, but fixed wavelength. 
Synthetic pulsar observations are generated by shifting the wavefield due
to the scintillation pattern in time and frequency in this manner, and by
choosing the length scales $r_{\rm diff}$ and $r_{\rm F}$ appropriate to
the scattering conditions.  The MCF is computed directly from the
wavefield, where the
wavenumbers, $k_\mr{min}$ and $k_\mr{max}$, were chosen to match a
5\,MHz bandwidth around 328\,MHz and the upper and lower cut-offs of
the power spectrum, $q_\mr{min}$ and $q_\mr{max}$, were set at values corresponding to
a physical size much larger than the scale of the phase screen and 
much smaller than the size of one discrete point, respectively. Table~\ref{tab:Scintillation Parameters} shows the
values of the other scintillation parameters with a brief explanation. The distance of the scintillation
screen, $D$, was chosen to be in the middle between the pulsar and the
observer. By tuning $Q_0$, we tried to create scintillation similar to that 
in our data from PSR B0329+54. The simulation results for PSR B0329+54 are shown in Fig.
~\ref{fig:SimScint} and clearly show an excess similar to that in our MCF of PSR B0329+54. 
This simulation does not reproduce the peak in the first bins. However, in order to see the 
effects of the short time scale variations of the scintillation pattern, we have simulated a 
growing scintillation pattern similar to that visible in Fig.~\ref{fig:Bandpass}b on top of a flat bandpass. 
Note that the individual modulations in the bandpass of our data from PSR B0329+54 often overlap each other.
This simulation did result in a peak in the first bins of the MCF. Furthermore, we split up our data
from PSR B0329+54 into smaller sets of data, calculated the MCF of each dataset and averaged them together.
We did this for datasets ten to hundred times shorter than the total length of our data and saw that 
the peak in the first bins of the MCF decreases when the dataset becomes smaller.

\begin{table}[!ht]
\flushleft
\caption{List of scintillation parameters and values used in the simulation}
 \begin{tabular}{ll}
  \hline
  Input & Discription \\
  parameters & \\
  \hline
  $k_\mr{min} = 6.807\,\mr{m^{-1}}$& Minimum wavenumber \\
  $k_\mr{max} = 6.912\,\mr{m^{-1}}$& Maximum wavenumber \\
		& ($k_\mr{min}$ and $k_\mr{max}$ are the edges of\\
		& the observed frequency range) \\
  $D = 0.7$\,kpc& Distance between scintillation \\
	& screen and observer\\
  $Q_0$ $= 1.1\times10^{-41}\,\mr{m}^{-17/3}$ & Amplitude of the power
  spectrum\\
        & of the phase fluctuations \\
  $N = 128$& Number of discrete points across \\
	& the wavefield (Total number of\\
      &  grid points = $N^2$)\\
  \hline
  Measured & \\
  parameters & \\
  \hline
  $r_\mr{F} = 1.7\times10^{9}$\,m& Fresnel scale ($=\sqrt{D/k}$) \\
  $r_\mr{diff} = 3\times10^{7}$\,m & Diffraction length scale \\
  $r_\mr{ref} = 1.7\times10^{11}$\,m & Refraction length scale \\
  & (= ${r_\mr{F}}^2/r_\mr{diff}$)\\
  $\Delta \nu_{\rm dc} = 0.02$\,MHz & Diffractive decorrelation \\
  & bandwidth \\
  \hline
 \end{tabular}
\label{tab:Scintillation Parameters}
\end{table}

\begin{table*}[!ht]
\flushleft
\caption{Diffractive decorrelation bandwidth and coherence
time of pulsars B0329+54, B0823+26, B0950+08, B1133+16 and
B1937+21.} 
\begin{tabular}{lccccc}
\hline
Pulsar & Frequency & Time resolution & $\Delta \nu_{\rm dc}$ & $1/\Delta \nu_{\rm dc}$ & Measured
coherence \\
  & (MHz) & (ns) &       & ($\mu$s) & time ($\mu$s)\\
\hline
B0329+54 & 328 & 100 & 29\,kHz~[1] &34 &35 [4]\\
B0823+26 & 430 & 100 & $0.81$\,MHz~[1] & 1.2 & 1.5 [5] \\ 
B0950+08 & 382 & 50  & $0.22 \pm 0.03$\,GHz~[3] &4.5$\times10^{-3}$ & none [4]\\
B0950+08 & 430 & 100 & $0.35 \pm 0.05$\,GHz~[3]  & 2.9$\times10^{-3}$ & 0.4 [5] \\
B1133+16 & 430 & 100 & $1.47$\,MHz~[1] & 0.67 & 1.1 [5]  \\
B1937+21 & 430 & 100 & 4.2 $\pm$ 0.9 kHz~[2]    & 238 & none [5] \\
\hline
\end{tabular}\\
\begin{tabular}{rl}
  \footnotesize
      [1] & \citet{Cordes86} \\{}
      [2] & \citet{Cordes90} \\{}
      [3] & \citet{Phillips92} \\{}
      [4] & Our data \\{}
      [5] & Data from JAP \\
      \hline
\end{tabular}
\normalsize
\label{tab:ScintDec}
\end{table*}

\section{Discussion and Conclusions}
\label{sec:discussion}

We find that frequency modulation
due to scintillation and possibly the shape of the WSRT bandpass have an effect on the MCF at small delay
values. This may be problematic in determining a coherence time.
For PSR B0950+08 we find no signature of coherence. For PSR B0329+54 we find
two features at small delay values: a peak in the first two bins and a broad excess reaching
up to 35\,$\mu$s. 
We have shown that the shape of the bandpass can
theoretically
cause a peak in the first bins of the MCF. The bandpass correction,
which uses the off-pulse, might not be effective in the case of PSR
B0329+54 due to the frequency modulation present in the on-pulse.
By comparing Fig.~\ref{fig:SimScint} with
Fig.~\ref{fig:OurResults1}b, we see that scintillation reproduces the
broad rise observed in the MCF of PSR B0329+54.
Furthermore, a simulation of an increasing
scintillation pattern on top of a flat bandpass, where the modulation due to the scintillation 
was overlapping, has shown that small changes in the scintillation pattern can cause 
a peak in the first bins of the MCF. This is similar to a time dependent
variation in the bandpass itself. Moreover, when smaller timescales are used to calculate the
MCF the peak is seen to decrease also indicating that scintillation may play a role here.

From the above, we conclude that scintillation can be responsible for
both the broad excess as well as the peak in the first two bins 
of the MCF of PSR B0329+54. 

We now discuss whether the excess found in the MCF of three
pulsars by JAP could also be due to scintillation. 
In Table~\ref{tab:ScintDec} we show the similarity between the
diffractive decorrelation bandwidth of the different pulsars and their
coherence time, defined
as the point where the MCF becomes zero. The values for $\Delta \nu_{\rm
dc}$ were scaled from a frequency of 51\,MHz for PSR B0950+08 and from
1\,GHz for the other pulsars, assuming $\Delta \nu_{\rm
dc}\ \propto \nu^{4.4}$~\citep{Cordes85}. No errors were quoted
in~\citet{Cordes86}, however it is known that $\Delta\nu_{dc}$ can vary significantly with time, in
some cases as much as a factor of 2$-$3~\citep{Bhat99}.
Assuming that the rise in the
MCF is due to
scintillation, we estimate the MCF to become zero when $\Delta\phi$ is on
the order of $1/\Delta\nu_\mr{dc}$. Looking at
Table~\ref{tab:ScintDec}, we see that the values of the
reciprocal of the diffractive decorrelation bandwidth and the measured
coherence time (fourth and fifth column, respectively) are indeed
similar. For our result of PSR B0950+08 there is no measured
coherence time, while JAP find a coherence time for
this pulsar of 0.4\,$\mu$s. Assuming the diffractive decorrelation
bandwidth from~\cite{Phillips92} the time resolution of our data and
that of JAP are too large to see the scintillation. The observed
feature in the MCF for PSR B0950+08 of JAP (see their Fig. 1) might then
result from a time variation in the bandpass of their data. There is,
however, some controversy as to what
  the diffractive decorrelation bandwidth for PSR B0950+08 is. According
to~\cite{Cordes86} $\Delta\nu_\mr{dc}$ is 4.0\,MHz at an observing
frequency of 430\,MHz. This gives a value for $1/\Delta\nu_\mr{dc}$ of
0.25\,$\mu$s, which is in the order of size of the coherence time of
0.4\,$\mu$s, measured by JAP. Assuming the value given by~\cite{Cordes86},
it would appear that the signal-to-noise ratio in our data is
insufficient to see the scintillation, as we see no excess in the MCF.
For PSR B1937+21, JAP also found no coherence time. Although this
pulsar is known to scintillate, the value for $1/\Delta\nu_\mr{dc}$ is so large that we can expect
the excess in the MCF due to scintillation to be smeared out over such
a large range of delays as to make it unmeasurable with the sensitivity of their observation.

We conclude that scintillation and possibly the shape of the bandpass causes the
excess in the MCF of our data from PSR B0329+54. Furthermore, we conclude that the coherent features, found
by JAP, also appear to be the result of scintillation. We therefore
cannot confirm that the MCF is clearly showing us the presence of coherence in these pulsars.

The authors would like to thank J. Cordes for his extensive comments and F. A. Jenet for his helpful discussions
which have both greatly contributed to the accuracy and clarity of this paper.

\bibliographystyle{apj}
\bibliography{H4303}

\clearpage

\end{document}